# Theoretical and experimental investigation of vacancy-based doping of monolayer MoS$_2$ on oxide.

*Amithraj Valsaraj,*[*,†] *Jiwon Chang,*[†] *Amritesh Rai,*[†] *Leonard F. Register,*[†] *and Sanjay K. Banerjee*[†]

[†]Microelectronics Research Center and Department of Electrical and Computer Engineering, The University of Texas at Austin, Austin, TX-78758, USA and [†]SEMATECH, 257 Fuller Road #2200, Albany, NY-12203, USA.

**ABSTRACT:** Monolayer transition metal dichalcogenides are novel, gapped two-dimensional materials with unique electrical and optical properties. Toward device applications, we consider MoS$_2$ layers on dielectrics, in particular in this work, the effect of vacancies on the electronic structure. In density-functional based simulations, we consider the effects of near-interface O vacancies in the oxide slab, and Mo or S vacancies in the MoS$_2$ layer. Band structures and atom-projected densities of states for each system and with differing oxide terminations were calculated, as well as those for the defect-free MoS$_2$-dielectrics system and for isolated dielectric layers for reference. Among our results, we find that with O vacancies, both the Hf-terminated HfO$_2$-MoS$_2$ system, and the O-terminated and H-passivated Al$_2$O$_3$-MoS$_2$ systems appear metallic due to doping of the oxide slab followed by electron transfer into the MoS$_2$, in manner analogous to modulation doping. The n-type doping of monolayer MoS$_2$ by high-$k$ oxides with oxygen vacancies then is experimentally demonstrated by electrically and spectroscopically characterizing back-gated monolayer MoS$_2$ field effect transistors encapsulated by oxygen deficient alumina and hafnia.

**INTRODUCTION**

The unique electrical and optical properties of monolayer (ML) transition metal dichalcogenides (TMDs) [1,2] have spurred intense research interest towards development of nanoelectronic devices utilizing these novel materials [3–10]. The atomically thin form of ML TMDs translates to excellent electrostatic gate control even at nanoscale channel length dimensions [11–13]. However, the two dimensional (2D) nature of ML TMDs makes their properties susceptible to the surrounding environment, as evidenced by the mobility enhancement of ML MoS$_2$ when placed on a high-$k$ dielectric such as hafnia (HfO$_2$) [4]. This mobility improvement in 2D materials was attributed to the damping of Columbic impurity scattering by high-$k$ dielectrics [14]. Theoretical calculations of HfO$_2$ interfaces have indicated that band offsets can be altered chemically by utilizing different interface terminations [15]. The conductive characteristics of MoS$_2$ deposited on SiO$_2$ have been shown to be dependent on the interface structure [16]. Controllable n-type doping of graphene transistors with extended air stability have been demonstrated by using self-encapsulated doping layers of titanium sub-oxide (TiO$_x$) thin films [17]. These results puts into stark focus the need to consider the effect of surrounding materials and the interfaces with them on the characteristics of ML TMDs.

In an earlier preliminary study [18], we considered MoS$_2$ on O- and Hf (Al)- terminated HfO$_2$ (Al$_2$O$_3$) via density functional theory (DFT). Those results suggest that O-terminated and H-passivated HfO$_2$ and Al$_2$O$_3$ exhibit potential as good substrates or gate insulators for ML MoS$_2$, with a straddling gap (Type 1) band structure for the composite system devoid of any defect states within the band gap of the MoS$_2$. However, ML MoS$_2$ on Hf-terminated HfO$_2$ shows a staggered gap alignment (Type 2), such that holes would be localized in the oxide rather than the MoS$_2$ layer. But in the case of ML MoS$_2$ on Al-terminated Al$_2$O$_3$ slab, we found that there is a straddling gap (Type 1) alignment to again provide localization of holes to the MoS$_2$ layer, although with some spill over into the nearby surface O states. We also considered O vacancies for MoS$_2$ on O terminated slabs of HfO$_2$ and Al$_2$O$_3$, observing significant doping in the latter. However, we did not include the van der Waal's interactions between the MoS$_2$ and oxide as we do in this work, tease out doping mechanism, consider the remaining combinations of oxide terminations, consider vacancies in the MoS$_2$, nor provide any experimental results.

In this work, we focus on the effects of O vacancies (O deficiency) in MoS$_2$ on HfO$_2$ and on Al$_2$O$_3$, and the effects of Mo and S vacancies in MoS$_2$ on HfO$_2$. We have used both density functional theory (DFT) and experimental analysis. For the O deficient systems, two possible terminations for the HfO$_2$ (Al$_2$O$_3$) slab are considered using DFT: an O-terminated HfO$_2$ (Al$_2$O$_3$) slab with H passivation and an Hf (Al)-terminated HfO$_2$ (Al$_2$O$_3$). The naming of two possible terminations is indicative of the initial structures used as starting point in our atomistic relaxations. The effects of O-vacancies in the first few layers of oxide on the band structure of the MoS$_2$-oxide system were simulated, with results for vacancies in the topmost/MoS$_2$-adjacent O layer shown here. Among our findings, O vacancies can lead to modulation-like doping of the MoS$_2$ from donor states in the oxide depending on the oxide terminations. Moreover, consistent with our theoretical results, electron doping of ML MoS$_2$ via O deficiency in the high-$k$ oxides was experimentally demonstrated by electrically and spectroscopically characterizing back-gated ML MoS$_2$ field effect transistors (FETs) encapsulated by O deficient alumina (Al$_2$O$_x$) or O deficient hafnia (HfO$_x$).

**METHODS**

**Computational Details**

The DFT calculations were performed using the projector-augmented wave method with a plane-wave basis set as implemented in the Vienna *ab initio* simulation package (VASP) [19,20]. We chose a kinetic energy cutoff of 400 eV. The k-mesh grid of 7x7x1 for the sampling of the first Brillouin zone of the supercell was selected according to Monkhorst-Pack type meshes with the origin being at the Γ point for all calculations except the band structure calculation. The local density approximation (LDA) [21] was employed primarily for the exchange-correlation potential as LDA has been shown to reproduce the apparent experimental band gap ($E_g$ =1.8 eV) [1] of ML MoS$_2$ well [22,23]. The calculated lattice constant for the MoS$_2$ layer after volume relaxation, $a$ = 3.122 Å, is also a good match to the experimental value [24]. We have also re-checked some of the DFT results using the generalized gradient approximation (GGA) [25]. We note, however, that both the LDA and the GGA underestimate the band gap of at least the bulk HfO$_2$ and Al$_2$O$_3$, which makes the prediction of band offsets from theoretical calculations unreliable. With approximately 150 atoms per supercell, use of presumably more accurate hybrid functionals or GW methods for atomistic relaxations was not practical. However, we have utilized hybrid functionals, namely HSE06 [26], to perform band structure calculations using the relaxed structures from our GGA simulations to further check our key conclusions. However, the primary objective of this theoretical work is to explore possible pathways to insulating and doping MoS$_2$ MLs qualitatively toward device applications, ultimately for experimental follow-up for promising cases. Similarly, we did not include spin orbit coupling here, which causes substantial spin splitting in the valence band, for similar reason. However, only conduction band doping is observed in our results, mitigating the impact of this latter approximation. Van der Waal's forces also were simulated due to the absence of covalent bonding between the TMD and the oxides [27]. In our computations, we have adopted the DFT-D2 scheme to model the non-local dispersive forces wherein a semi-empirical correction is added to the conventional Kohn-Sham DFT theory [28].

The two representative dielectrics, HfO$_2$ and Al$_2$O$_3$, were chosen for high-k value and minimal lattice mismatch, respectively. The MoS$_2$ ML of principle interest, with its hexagonal lattice, was taken to be unstrained with its above-noted volume-relaxed lattice constant of $a$ = 3.122 Å. For the dielectric oxide, the energetically stable crystalline phases of bulk HfO$_2$ and Al$_2$O$_3$ at ambient conditions, namely, monoclinic HfO$_2$ [29] and hexagonal Al$_2$O$_3$ [30], respectively, were utilized. Our simulations were performed by constructing a supercell of ML MoS$_2$ on an approximately 2 nm thick oxide slab. For HfO$_2$, atomic relaxation was performed within a rectangular supercell ($a$ = 9.366 Å, $b$ = 5.407 Å) chosen to reduce the lattice mismatch between ML MoS$_2$ and monoclinic HfO$_2$. However, a roughly 6% strain remains along the in-plane directions in the HfO$_2$ (Figure 1(a)). For Al$_2$O$_3$, atomic relaxation was performed in a (rotated) hexagonal supercell ($a$ = 8.260 Å) with a strain of only about 0.2% (Figure 1(b)). The systems were relaxed until the Hellmann-Feynman forces on the atoms were less than 0.02 eV/Å. During relaxation, all the MoS$_2$ ML atoms and the top half of the layers of the dielectric oxide were allowed to move in all three spatial dimensions. Oxygen vacancies were modeled by removing a single O atom from an O-layer of the supercell. Since we have periodic supercells, the O vacancy is repeated in each instance of the supercell. The system is then allowed to relax again with the introduced O-vacancy. A similar procedure was followed in the modeling of Mo and S vacancies in the $MoS_2$-$HfO_2$ system. In the latter case, the S atom vacancy was introduced in the layer adjacent to the oxide surface. All simulations were performed at a temperature of 0 K.

**Experiment**

ML $MoS_2$ was mechanically exfoliated from commercially available bulk $MoS_2$ crystals (SPI Supplies) onto a degenerately doped n-type Si-(100) substrate, which served as the back-gate, covered by a 90 nm thick thermal oxide. Upon exfoliation, the samples were annealed at $350^0$ C in high vacuum (~ $10^{-6}$ Torr) for 8 hours to minimize tape residues and trapped adsorbates between the $MoS_2$ and the silicon dioxide substrate. A combination of optical microscopy, atomic force microscopy, Raman and photoluminescence measurements were used to identify atomically flat ML $MoS_2$ flakes of interest. Source and drain contacts were patterned using electron beam lithography followed by electron beam evaporation and solvent lift-off of an Ag/Au (20/30 nm) stack in the case of hafnia, or just Au (50 nm) in the case of alumina. Finally, devices were covered by ~ 30 nm of either alumina or hafnia deposited at $200^0$ C using atomic layer deposition (ALD) via the reaction of water with standard ALD precursors, namely trimethyl aluminum for alumina and tetrakis (dimethylamido) hafnium for hafnia (Figure 1(c)). An in-depth look at the experimental details and device fabrication procedures can be found in the work by Rai et al. [31,32]. The stoichiometry of the as-deposited high-k oxide was determined using x-ray photoelectron spectroscopy (XPS).

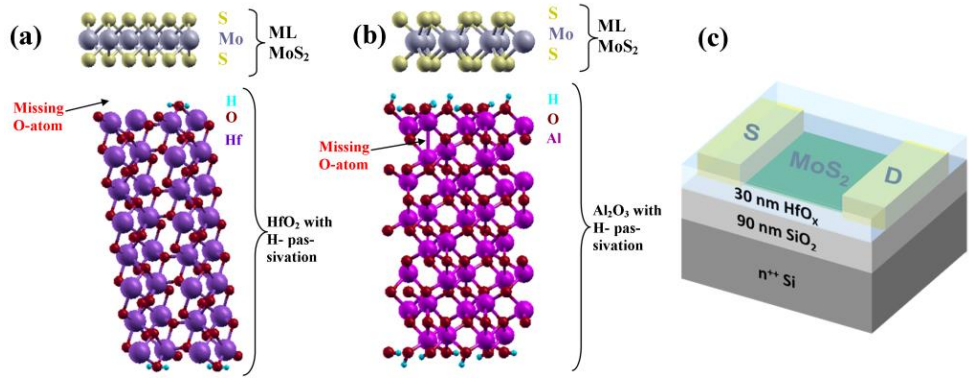

**Figure 1.** (a) Supercell of ML MoS$_2$ on an H-passivated, O-terminated HfO$_2$ slab of approximately 2 nm thickness with O-vacancy (side view). (b) Supercell of ML MoS$_2$ on an H-passivated, O-terminated Al$_2$O$_3$ slab of approximately 2 nm thickness with O-vacancy (side view). The monolayer of MoS$_2$ belongs to the space group P-6m2 (point group D$_{3h}$). (c) 3D schematic of a back-gated ML MoS$_2$ FET encapsulated by HfO$_x$.

**RESULTS AND DISCUSSION**

The band structure and atom-projected density of states (AP-DOS) have been calculated for the ML MoS$_2$-oxide system considering different possible terminations of the oxide at the interface in the presence of O vacancies in the oxide or Mo and S vacancies in the MoS$_2$. We compared (overlaid) the band structures for the MoS$_2$-oxide systems with vacancies to the ideal MoS$_2$-oxide results. In all cases, the highest occupied state of the system with vacancies serves as the zero energy reference in these 0 K simulations. However, the reference band structures absent vacancies are shifted up or down to provide a rough fit to the former in terms of band structure and the atom projected densities of states (AP-DOS) of the Mo and S atoms. (Otherwise, the zero energy reference for the latter would be the valence band edge.)

**Monolayer MoS$_2$ on HfO$_2$ slab with O vacancy**

When an O vacancy is introduced into the top layer of the O-terminated and H-passivated HfO$_2$ slab, in these 0 K simulations, an occupied defect state (band) is introduced within the band gap of ML MoS$_2$ (Figure 2(a)), which is associated primarily with Hf atoms in the oxide.

Analogous Hf-associated defect states also arise in an isolated O-terminated and H-passivated HfO$_2$ slab (Figure 3(a) and (b)). In this latter case (and for analogous cases below) we simply removed the MoS$_2$ layer from the combined system, while otherwise holding the crystal structure fixed as a control. However, the close proximity of the occupied defect band to the conduction band (of the reference band structure) suggests that these states might be able to act as donors. As can be seen from the AP-DOS (Figure 2(b)), the conduction band edge for MoS$_2$ is pinned at the Fermi level indicating n-type doping. However, the defect band formation due to the limited supercell size and associated very large ($1.97 \times 10^{14}$/cm$^2$) O-vacancy density in these simulations leaves the binding energy for lower defect densities uncertain. Alternatively, these interface states could function as relatively shallow charge traps, leading to degradation of device performance. Since a rectangular supercell was used in these simulations of MoS$_2$ on HfO$_2$, the corresponding Brillouin zone (BZ) is smaller and the K point of the primitive unit cell—where the ML MoS$_2$ band edges are located—folds into the Γ point in the supercell's BZ.

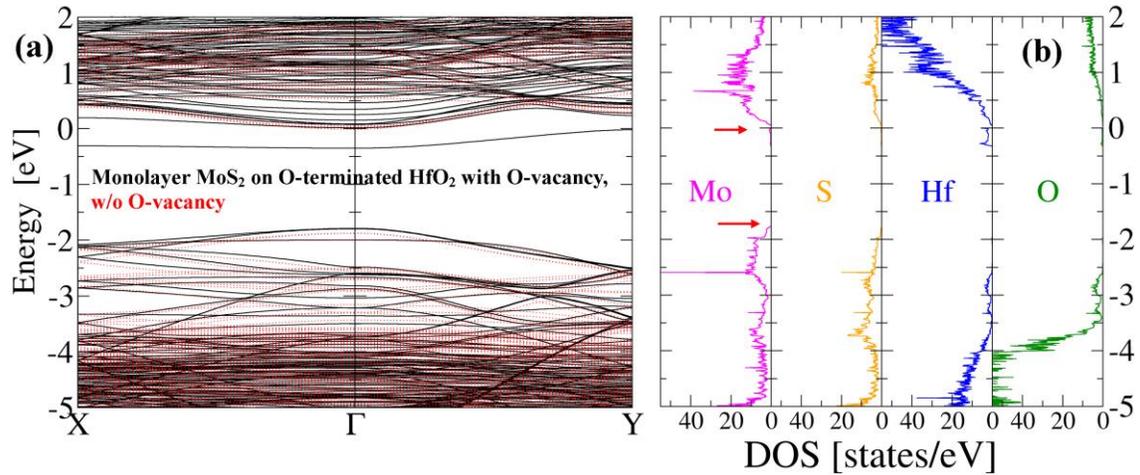

**Figure 2**. (a) Band structure of ML MoS$_2$ on an H-passivated, O-terminated HfO$_2$ slab with an O-vacancy in the top layer, plotted along the high symmetry directions of the BZ (black solid lines). The 0 eV reference corresponds to the highest occupied state in these 0 K simulations. The band structure of vacancy-free ML MoS$_2$-HfO$_2$ system (O-terminated) is superimposed for comparison (red dashed lines). However, this latter band structure, which otherwise would have its zero reference energy at the upper edge of the valence band, is shifted up or down to provide a reasonable fit to the former (b) Atom-projected density of states for the ML MoS$_2$ and O-terminated HfO$_2$ system with an O-vacancy. Red arrows indicate the conduction and valence band edges. An occupied defect state (band) is introduced within the band gap of ML MoS$_2$.

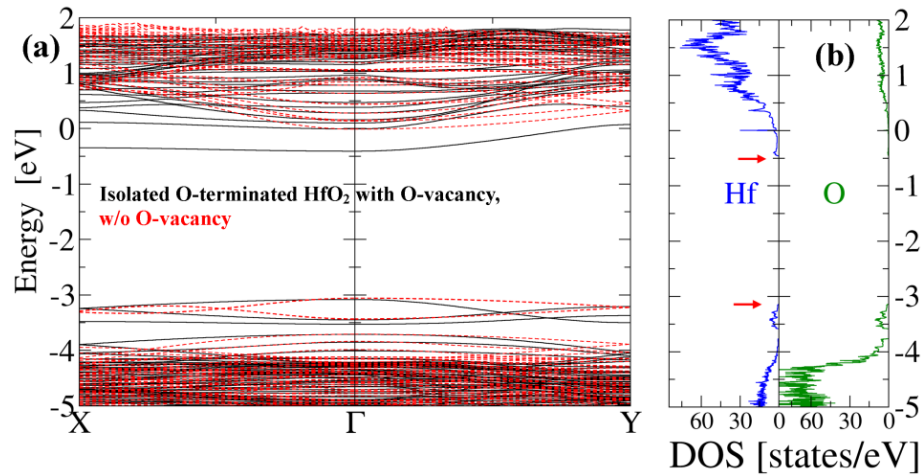

**Figure 3**. (a) Band structure of a freestanding H-passivated, O-terminated HfO$_2$ slab with an O-vacancy in the top layer, plotted along the high symmetry directions of the BZ (black solid lines). The energy-shifted band structure of vacancy-free HfO$_2$ slab (O-terminated) is superimposed for comparison (red dashed lines). (b) Atom-projected density of states for the O-terminated HfO$_2$ system with an O-vacancy. Defect states associated with Hf-atoms are observed.

In the case of Hf-terminated $HfO_2$-$MoS_2$ system with an O vacancy in the top layer of oxide, there is a straddling gap alignment (Type-1) as seen in the AP-DOS (Figure 4(b)) for this large O-vacancy density, much as for O-terminated $HfO_2$. Moreover, there are now two partially occupied bands at the bottom of the conduction band (Figure 4(a)), both of which are largely localized to the $MoS_2$ layer, resulting in a system that now appears metallic. Calculation of the band structure for a freestanding Hf-terminated $HfO_2$ slab with an O vacancy exhibits occupied conduction band states associated with the Hf atoms (Figure 5(a) and (b)). In the combined $HfO_2$-$MoS_2$ system, these electrons are then transferred into the lower conduction-band-edge $MoS_2$ layer, in a modulation-doping-like process. In $MoS_2$, the DOS at the conduction and valence band edges are dominated by $d_{xz}$ and $d_{z^2}$ orbitals from the Mo atoms while in the $HfO_2$ the band edge states arise mainly from the contribution of Hf- $d$ orbitals and O- $p$ orbitals.

For the $HfO_2$-$MoS_2$ with O vacancy systems, we also repeated the simulations with the GGA approximation for comparison with the above LDA results. Figure 6(a) shows the band structure of ML $MoS_2$ on Hf-terminated $HfO_2$ with an O-vacancy, as obtained using both the GGA and the LDA approximations. The same nominal crystal structure was used, but a separate relaxation was performed for the LDA and GGA calculations (the latter, however, starting with the former for computational efficiency). As can be seen, the results match closely, including the degree of degenerate doping. A similar comparison (not shown) was performed for $MoS_2$ on O-terminated $HfO_2$, again with good agreement between the results obtained with the GGA and with the LDA including the location of the occupied defect band just below the conduction band. Finally, in Figure 6(b), we have used hybrid functionals, specifically HSE06, which provide a more accurate value for the band gap of bulk $HfO_2$ to simulate the band structure of ML $MoS_2$ on Hf-terminated $HfO_2$, to further check key results. The much larger computational demands required for hybrid method combined with the large supercell size constrained us to use a coarse k-point grid for evaluation of the band structure and precluded us from running any relaxations of the structure using the hybrid method. Instead, we reused the structure obtained from the GGA relaxations. As shown in Figure 6(b), with the hybrid method, the conduction band edge is again pulled below the Fermi level as in our previous GGA and LDA results, indicating the n-type doping of ML $MoS_2$ modulated by dielectric oxide.

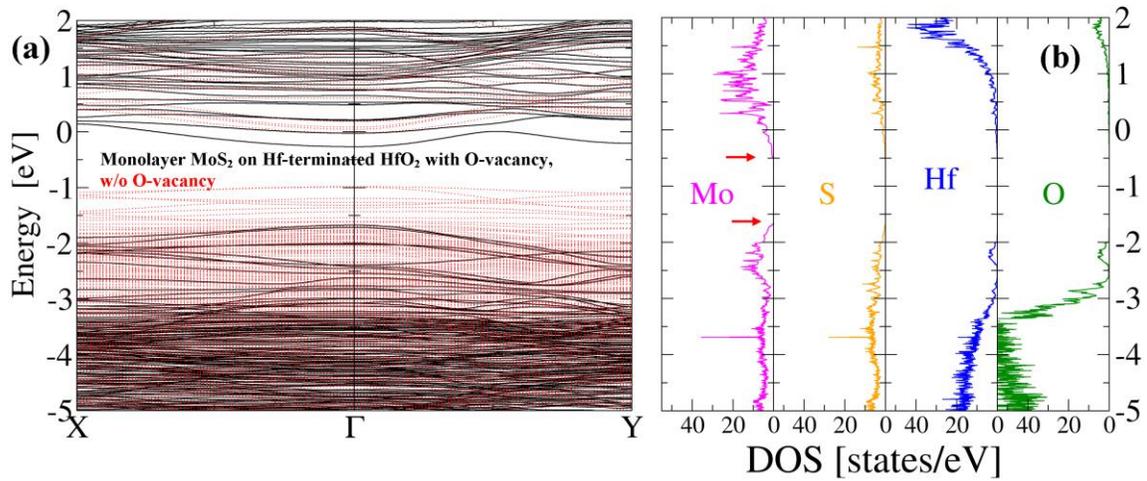

**Figure 4**. (a) Band structure of ML $MoS_2$ on Hf-terminated $HfO_2$ slab with an O-vacancy in the top layer, plotted along the high symmetry directions of the BZ (black solid lines). The energy-shifted band structure of vacancy free ML $MoS_2$-$HfO_2$ system with Hf-termination is superimposed for comparison (red dashed lines). (b) Atom-projected density of states for the ML $MoS_2$ and Hf-terminated $HfO_2$ system with an O-vacancy. A straddling gap band alignment is now observed along with two partially occupied bands at the conduction band edge both of which are largely localized to the $MoS_2$ layer, resulting in a system that now appears metallic.

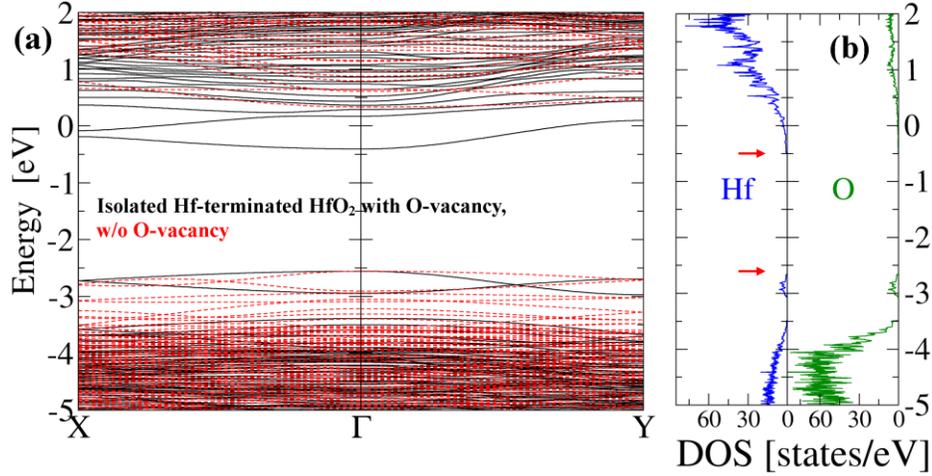

**Figure 5.** (a) Band structure of a freestanding Hf-terminated $HfO_2$ slab with an O-vacancy in the top layer, plotted along the high symmetry directions of the BZ. The energy-shifted band structure of vacancy-free $HfO_2$ slab (Hf-terminated) is superimposed for comparison (red dashed lines). (b) Atom-projected density of states for the Hf-terminated $HfO_2$ system with an O-vacancy. An occupied conduction band edge state associated with the Hf atoms is observed.

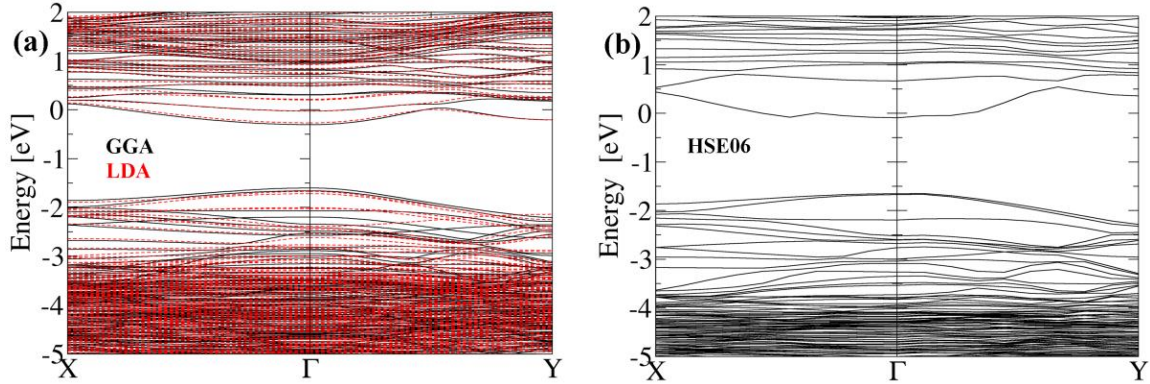

**Figure 6.** (a) Band structure of ML $MoS_2$ on an Hf-terminated $HfO_2$ with an O vacancy obtained using the GGA (solid lines, black online). The band structure obtained using the LDA is overlaid on top for comparison (dashed lines, red online). Both results exhibit n-type doping, and essentially the same degree of degeneracy. (The zero energy reference remains the Fermi level in each case). (b) Band structure of ML $MoS_2$ on an Hf-terminated $HfO_2$ with an O vacancy obtained using the HSE06. The conduction band edge is pulled below the Fermi level indicating n-type doping of $MoS_2$.

**Monolayer MoS$_2$ on Al$_2$O$_3$ slab with O vacancy**

For the O-terminated and H-passivated Al$_2$O$_3$-MoS$_2$ system, creation of an O vacancy in the top O-layer of Al$_2$O$_3$ produces only a modest effect on the conduction band edge states in comparison to the vacancy free reference system. However, the O-vacancy pulls the conduction band edge below the Fermi level, filling the lower MoS$_2$ conduction band states (Figure 7(a)), which remain largely localized in space to the MoS$_2$ layer (Figure 7(b)), resulting in a system that now appears metallic, much as for the Hf-terminated HfO$_2$-MoS$_2$ system with an O vacancy. Calculation of the band structure for an isolated O-terminated Al$_2$O$_3$ slab with an O vacancy exhibits occupied conduction band states associated with the O atoms (Figure 8(a) and (b)). In the combined Al$_2$O$_3$-MoS$_2$ system, these electrons again are transferred into the lower conduction-band-edge MoS$_2$ layer, in a modulation-doping-like process.

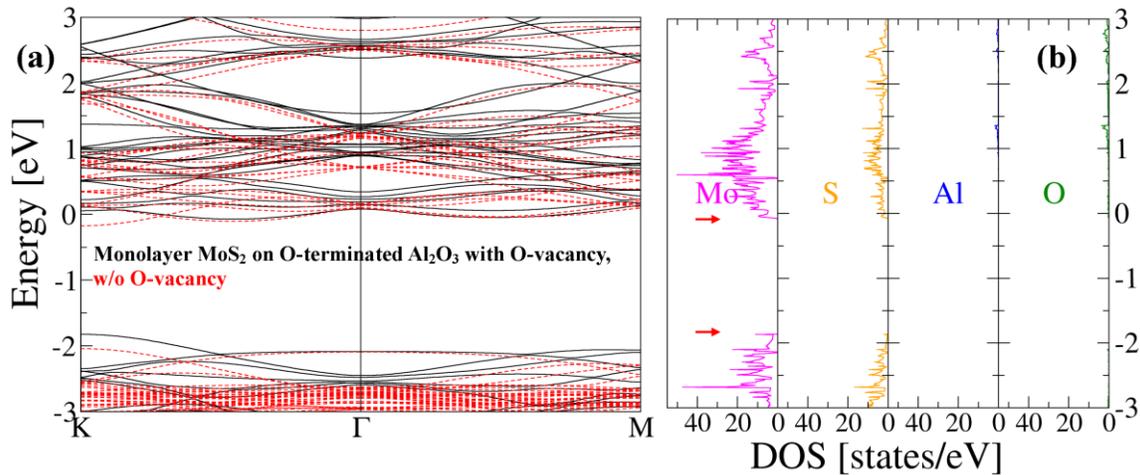

**Figure 7.** (a) Band structure of ML MoS$_2$ on an H-passivated, O-terminated Al$_2$O$_3$ slab with an O-vacancy in the top layer, plotted along the high symmetry directions of the BZ (black solid lines). The energy-shifted band structure of vacancy free ML MoS$_2$-Al$_2$O$_3$ system (O-terminated) is superimposed for comparison (red dashed lines). (b) Atom-projected density of states for the ML MoS$_2$ and O-terminated Al$_2$O$_3$ system with an O- vacancy. A new partially filled band is introduced at the edge of the MoS$_2$ conduction band, which is largely localized to the MoS$_2$ layer, resulting in a system that now appears metallic.

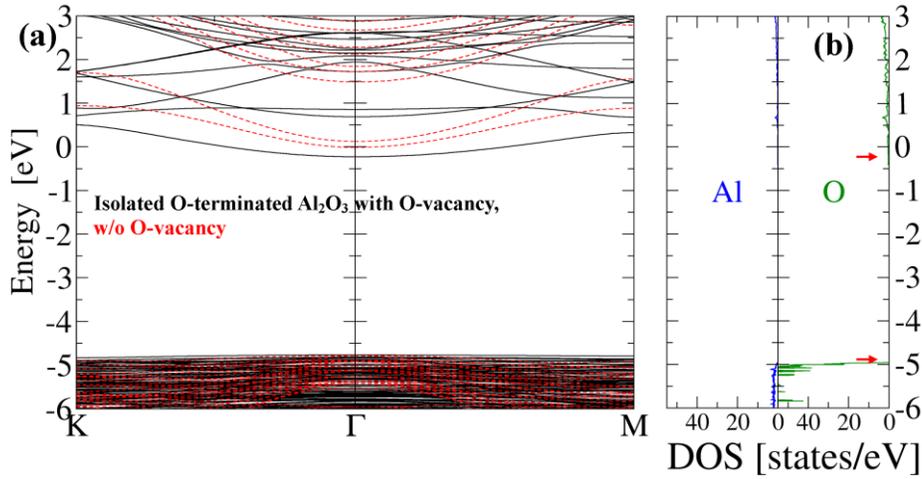

**Figure 8**. (a) Band structure of freestanding H-passivated, O-terminated $Al_2O_3$ slab with an O-vacancy in the top layer, plotted along the high symmetry directions of the BZ. The energy-shifted band structure of vacancy-free $Al_2O_3$ slab (O-terminated) is superimposed for comparison (red dashed lines). (b) Atom-projected density of states for the O-terminated $Al_2O_3$ system with an O-vacancy. An occupied conduction band edge state associated with the O atoms is observed.

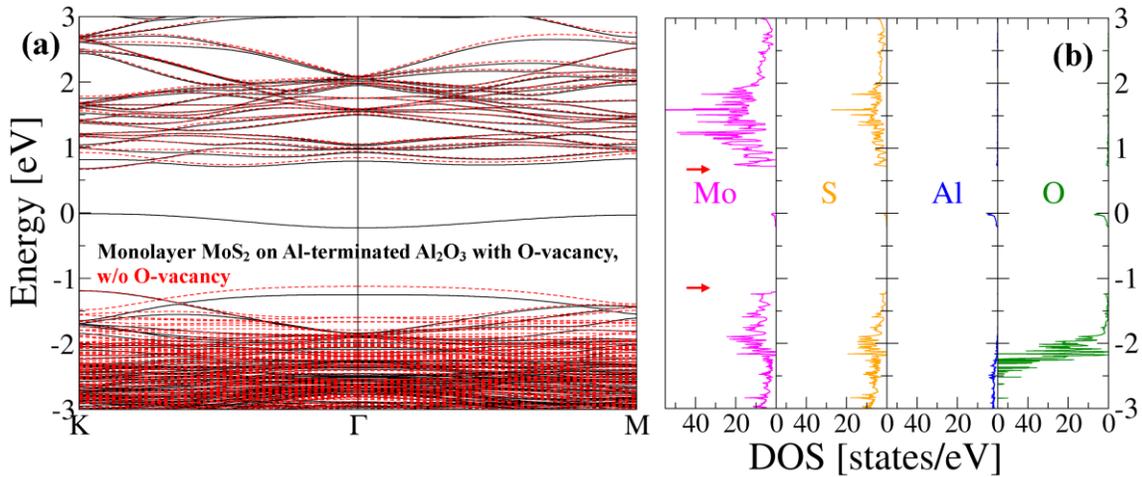

**Figure 9**. (a) Band structure of ML $MoS_2$ on Al-terminated $Al_2O_3$ slab with an O-vacancy in the top layer, plotted along the high symmetry directions of the BZ (black solid lines). The energy-shifted band structure of vacancy free ML $MoS_2$-$Al_2O_3$ system (Al-terminated) is superimposed for comparison (red dashed lines). (b) Atom-projected density of states for the ML $MoS_2$ and Al-terminated $Al_2O_3$ system with an O-vacancy. An occupied state (band) deep in the band gap of the $MoS_2$ is produced, which is localized to the Al and O atoms in the oxide layer.

For Al-terminated $Al_2O_3$-$MoS_2$ system, the system retains a straddling gap alignment after the introduction of an O vacancy in the oxide layer. However, an occupied state (band) deep in the band gap of the $MoS_2$ is produced (Figure 9(a)), which is localized to the Al and O atoms in the oxide layer (Figure 9(b)). Such defect states could serve as recombination centers or charge traps. In addition, however, a direct band gap is found at these doping concentrations, in contrast to the Al-terminated $Al_2O_3$-$MoS_2$ system without an O vacancy.

**Mo and S vacancies in $MoS_2$**

We introduced a single Mo or S atom vacancy in the supercell of the $MoS_2$ on O-terminated and H-passivated $HfO_2$ system, with the S atom vacancy introduced in the layer adjacent to the oxide surface. The corresponding vacancy density is $1.97 \times 10^{14}/cm^2$ in either case. The resultant band structure and AP-DOS are plotted in Figure 10 and Figure 11 for $MoS_2$-$HfO_2$ systems with Mo and S vacancies, respectively. In both cases, a straddling gap alignment is retained, but defect states are introduced into the band gap of the $MoS_2$ that are localized to the $MoS_2$ layer (Figs. 10(b) and 11(b)). With the Mo vacancy, several states are introduced within the nominal band gap, both occupied and empty, although at this vacancy concentration, the valence band edge is difficult to define (Figure 10(a)). In the case of S vacancies in the $MoS_2$ ML, there are two unoccupied defect states (bands) introduced near mid gap, as well as significant distortion of the valence band edge structure at these vacancy concentrations (Figure 11(a)).

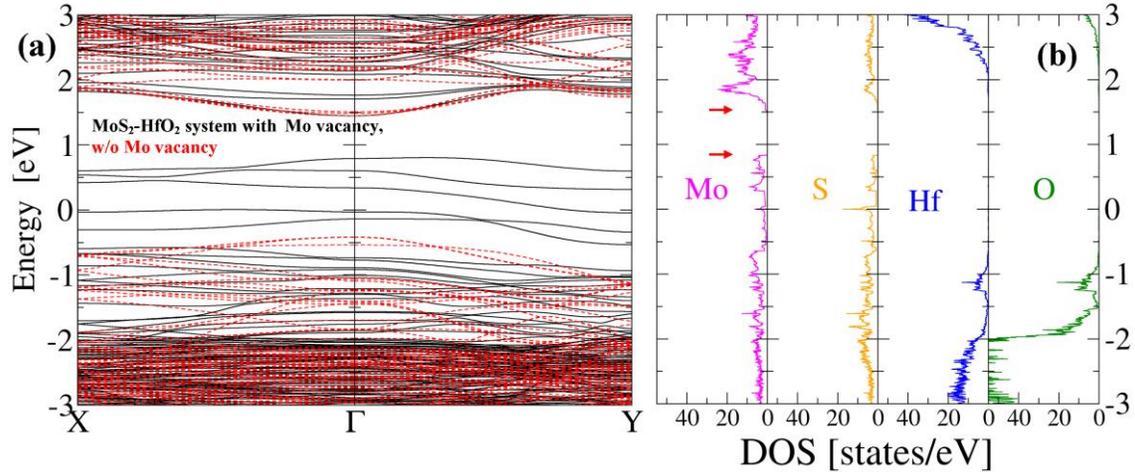

**Figure 10**. (a) Band structure of ML MoS$_2$ on an H-passivated, O-terminated HfO$_2$ slab with Mo-vacancy in the ML, plotted along the high symmetry directions of the BZ (black solid lines). The energy-shifted band structure of vacancy free ML MoS$_2$-HfO$_2$ system (O-terminated) is superimposed for comparison (red dashed lines). (b) Atom-projected density of states for the ML MoS$_2$ and O-terminated HfO$_2$ system with Mo vacancy. Several states are introduced within the nominal band gap and significant distortion of the valence band edge structure is observed.

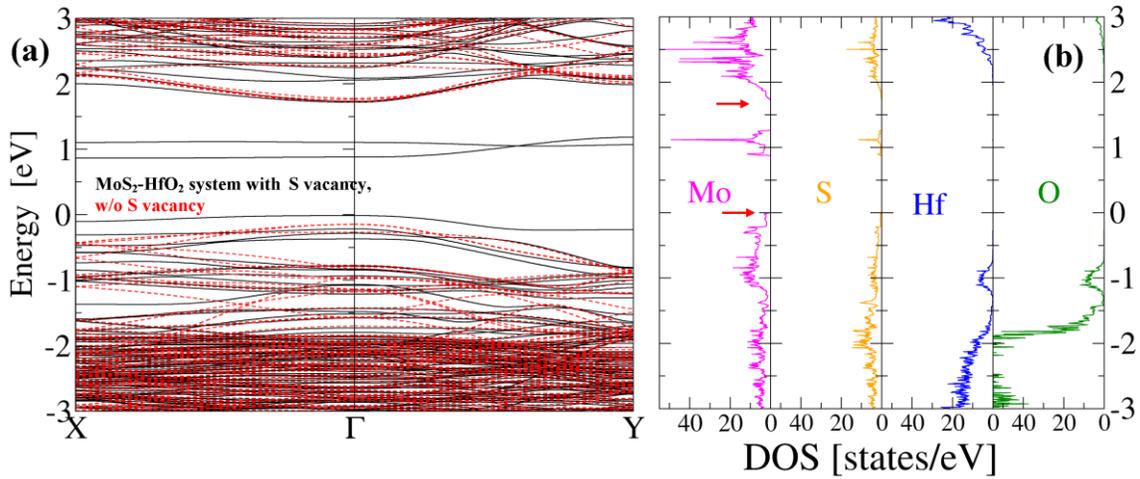

**Figure 11**. (a) Band structure of ML MoS$_2$ on an H-passivated, O-terminated HfO$_2$ slab with S-vacancy in the ML, plotted along the high symmetry directions of the BZ (black solid lines). The energy-shifted band structure of vacancy free ML MoS$_2$-HfO$_2$ system (O-terminated) is superimposed for comparison (red dashed lines). (b) Atom-projected density of states for the ML MoS$_2$ and O-terminated HfO$_2$ system with S vacancy. Two unoccupied defect states (bands) are introduced near mid gap, and significant distortion of the valence band edge structure is observed.

**Experimental Results**

Electron doping of ML MoS$_2$ by O deficient high-$k$ oxides was experimentally demonstrated by electrically and spectroscopically characterizing back-gated ML MoS$_2$ FETs encapsulated by alumina (Al$_2$O$_x$) and hafnia (HfO$_x$). The DFT calculations would suggest that an O-deficient high-$k$ oxide encapsulating the MoS$_2$ ML would produce a combination of n-type modulation doping of the bands and occupied defect states within the gap in bulk materials, the latter contributing perhaps little to the doping but important when trying to pull the Fermi level below them. On exposure to air, the Hf (Al)-terminated HfO$_2$ (Al$_2$O$_3$) is unrealistic while the O-termination provides a more accurate model for surface termination in the oxide. However, in our experimental system (Figure 1(c)), the high-$k$ oxide encloses the MoS$_2$ ML and MoS$_2$-oxide interface is not exposed to air allowing us to investigate both O-rich and O-deficient oxide interfaces.

Figure 12(a) shows the room temperature (RT) transfer characteristics of a back-gated ML MoS$_2$ FET before (blue) and after (red) encapsulation by ALD HfO$_x$. The length ($L_{CH}$) and width ($W$) of the device are 900 nm and 2 μm, respectively, and the data was collected at a drain-source voltage (V$_{DS}$) of 50 mV. Before encapsulation, the device exhibits a threshold voltage ($V_{th}$) near −15 to −20 V. After encapsulation in ALD HfO$_x$, there is a large negative shift in V$_{th}$ consistent with n-type doping, as well as pronounced stretch-out of the transfer characteristic as V$_{BG}$ is made more negative, consistent with near-band-edge defects in the band gap as predicted by the DFT for O-deficient HfO$_x$. The n-type doping was further confirmed by Raman spectroscopy performed on the ML MoS$_2$ in the channel region of the same FET, Figure 12(b), before (blue) and after (red) HfO$_x$ encapsulation. Before HfO$_x$, the peak positions of the out-of-plane A$_{1g}$ and the in-plane E$_{2g}^1$ peaks are at ~ 402 cm$^{-1}$ and ~ 383 cm$^{-1}$, respectively, which is characteristic of ML MoS$_2$ [33]. After HfO$_x$ encapsulation, the E$^1_{2g}$ peak remains relatively unchanged, while the A$_{1g}$ peak shows a distinct broadening and a red shift in its peak position from ~ 402 cm$^{-1}$ to ~ 399 cm$^{-1}$. These changes in the A$_{1g}$ Raman peak upon HfO$_x$ encapsulation are indicative of the increased electron concentration in the ML MoS$_2$ channel, and also have been observed in previous n-type doping studies of MoS$_2$ [34]. The Hf:O atomic ratio in the as-deposited HfO$_x$ was determined to be ~ 1:1.56 from XPS analysis, thereby establishing the correlation between oxygen deficiency and n-type doping of ML MoS$_2$ caused by HfO$_x$.

The results for the control sample of O-rich HfO$_x$ on ML MoS$_2$ are shown in Figure 13. The Hf:O ratio for the ALD deposited O-rich HfO$_x$ was determined to be ~ 1:2.1 from XPS measurements in exactly the same manner and using the same number of components that were used in peak fitting of the O-deficient HfO$_x$. As can be clearly seen, there is negligible change in the Raman spectra of MoS$_2$ after O-rich HfO$_x$ deposition. There is no red shift or peak broadening of the A$_{1g}$ Raman mode implying negligible n-type doping of MoS$_2$. Moreover, from the transfer curve we can see that the device can be turned off within the same back-gate voltage sweep range after deposition of the O-rich HfO$_x$. These results depict negligible doping of the ML MoS$_2$ after O-rich HfO$_x$ deposition, which contrast to the strong doping of the MoS$_2$ when an O-deficient HfO$_x$ was deposited.

Similar n-type doping results were obtained after encapsulating back-gated ML MoS$_2$ FETs with ALD Al$_2$O$_x$. Figure 14(a) and 14(b) show the RT transfer characteristics of a ML MoS$_2$ FET ($L_{CH}$ = 500 nm, $W$ = 2.3 µm, $V_{DS}$ = 100 mV, Au contacts) and the normalized Raman spectra of the ML MoS$_2$ channel, respectively, before (blue) and after (red) ALD Al$_2$O$_x$ deposition. As in the case of HfO$_x$, the negative $V_{th}$ shift (Figure 14(a)) and the broadening and red shift of the A$_{1g}$ Raman peak (Figure 14(b)) of ML MoS$_2$ after encapsulation in ALD Al$_2$O$_x$ is indicative of n-type doping, again consistent with the DFT results with oxygen vacancies. The Al:O atomic ratio was determined to be ~ 2:1.55 from XPS analysis, thereby, confirming the inherent oxygen deficiency in the as-deposited high-$k$ oxide.

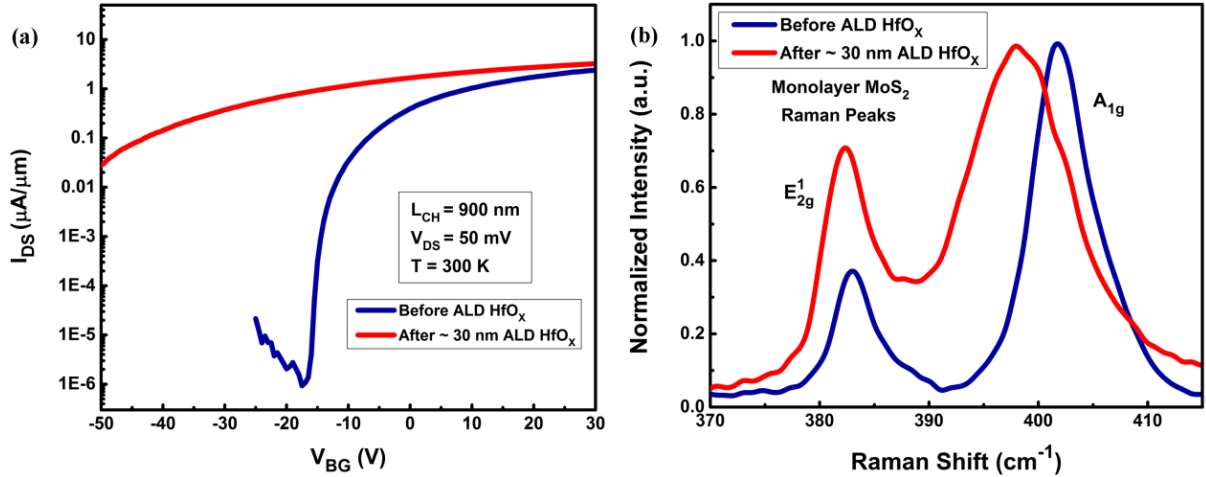

**Figure 12**. (a) Room temperature transfer characteristics of a back-gated ML MoS$_2$ FET before (blue) and after (red) ~ 30 nm ALD HfO$_x$ (x ~ 1.56) encapsulation, and (b) corresponding normalized Raman spectra of the ML MoS$_2$ FET channel before (blue) and after (red) ALD HfO$_x$ encapsulation. The shifted threshold voltage and A$_{1g}$ peak are consistent with n-type doping after encapsulation.

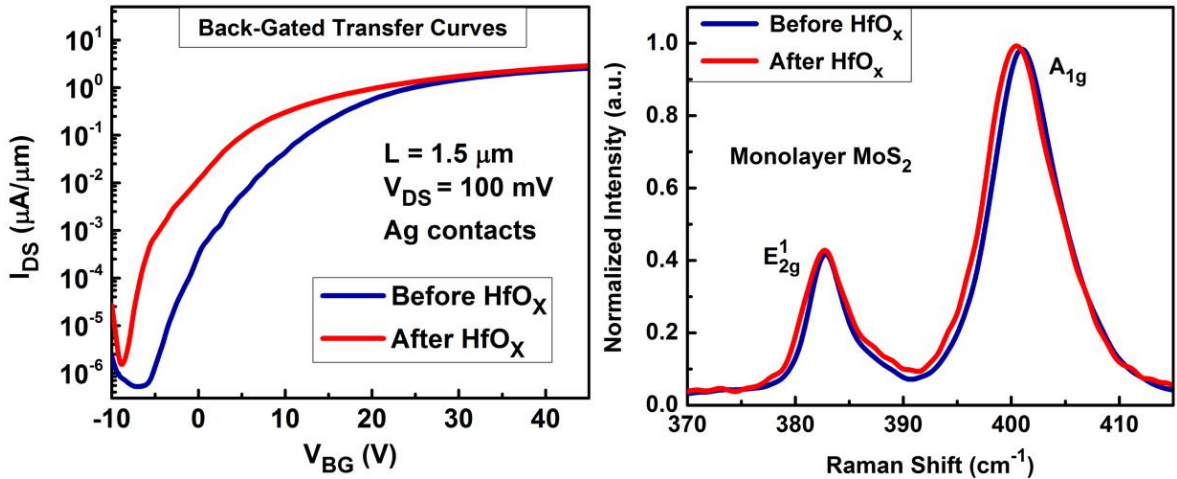

**Figure 13**. (a) Room temperature transfer characteristics of a back-gated ML MoS$_2$ FET before (blue) and after (red) ~ 30 nm ALD HfO$_x$ (x ~ 2.1) encapsulation, and (b) corresponding normalized Raman spectra of the ML MoS$_2$ FET channel before (blue) and after (red) ALD HfO$_x$ encapsulation. There is no red shift or peak broadening of the A$_{1g}$ Raman mode implying negligible n-type doping of MoS$_2$

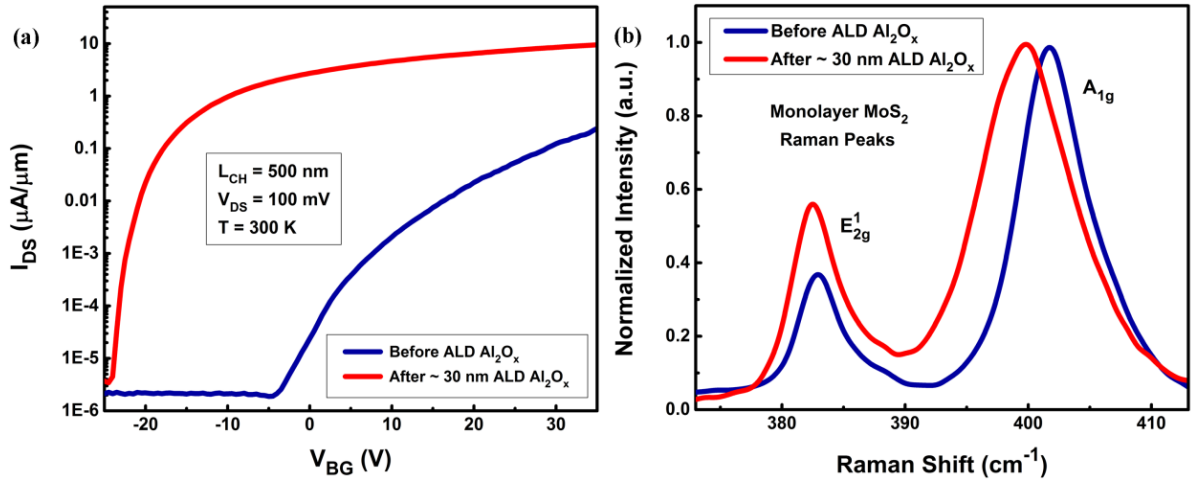

**Figure 14.** (a) Room temperature transfer characteristics of a back-gated ML MoS$_2$ FET before (blue) and after (red) ~ 30 nm ALD Al$_2$O$_x$ (x ~ 1.55) encapsulation and (b) corresponding normalized Raman spectra of the ML MoS$_2$ FET channel before (blue) and after (red) ALD Al$_2$O$_x$ encapsulation. The shifted threshold voltage and A$_{1g}$ peak are consistent with n-type doping after encapsulation.

**CONCLUSION**

In summary, DFT simulations suggest that occupied near-conduction-band-edge states that might function either as donors or shallow traps are introduced in the MoS$_2$-oxide system by O vacancies in the O-terminated and H-passivated HfO$_2$-MoS$_2$ system. More promising as a means of doping, with O vacancies, both the Hf-terminated HfO$_2$-MoS$_2$ system, and the O-terminated and H-passivated Al$_2$O$_3$-MoS$_2$ system appear metallic due to doping of the oxide slab followed by electron transfer into the MoS$_2$, in manner analogous to modulation doping. Consistent with these latter theoretical results, n-type doping of ML MoS$_2$ by high-*k* oxides with oxygen vacancies was demonstrated experimentally by electrically and spectroscopically characterizing back-gated ML MoS$_2$ FETs encapsulated by alumina (Al$_2$O$_x$) and hafnia (HfO$_x$). In contrast to the effects of vacancies in the oxides, in simulations, Mo and S vacancies in MoS$_2$ ML introduces multiple deep defect states in the band gap of MoS$_2$, as well as distorting the band edges.

AUTHOR INFORMATION

**Corresponding Author**

*Amithraj Valsaraj: amithrajv@utexas.edu

ACKNOWLEDGMENT

This work is supported by SEMATECH, the Nanoelectronics Research Initiative (NRI) through the Southwest Academy of Nanoelectronics (SWAN), and Intel. We thank the Texas Advanced Computing Center (TACC) for computational support.